**Associate Professor Bogdan MÂRZA, PhD**
**Assistant Professor Renate BRATU, PhD**
**Professor Răzvan ȘERBU, PhD**
**Faculty of Economic Sciences**
**"Lucian Blaga" University of Sibiu**
**Assistant Professor Sebastian STAN, PhD**
**Faculty of Military Management**
**"Nicolae Balcescu" Land Army Academy of Sibiu**
**Professor Camelia OPREAN-STAN, PhD (Corresponding author)**
**E-mail: camelia.oprean@ulbsibiu.ro**
**Faculty of Economic Sciences,**
**"Lucian Blaga" University of Sibiu**

# APPLYING AHP AND FUZZY AHP MANAGEMENT METHODS TO ASSESS THE LEVEL OF FINANCIAL AND DIGITAL INCLUSION

*Abstract. In today's world, marked by social distancing and lockdowns, the development of digital financial services is becoming increasingly important, but there is little empirical work documenting the most important factors that contribute to the process of financial and digital inclusion. Because the speed with which states adapt to digital financial services is critical, we must ask how prepared states are for this transition and how far they have progressed in terms of financial and digital inclusion. In this context, the goal of this article is, on the one hand, to propose a financial responsibility process framework capable of raising awareness of the most important harmonized key levels of financial and digital inclusion process that, when properly managed, can lead to achieving an optimal level of financial responsibility, and, on the other hand, to assess the financial and digital inclusion process of two different age groups of individuals who are active in the financial environment (15-34 and 35-59 age groups). The Analytical Hierarchy Process AHP and Fuzzy AHP approaches are proposed as a framework for assessing the mechanism of financial and digital inclusion in five East Central European countries. The findings reflect differences between the analyzed countries in terms of the key levels of financial and digital inclusion (where digital and financial education are the most important levels), with Croatia, Czech Republic, and Poland being the most integrated and Romania being the least. According to the findings, as a country or region's level of financial and digital inclusion increases, so does its level of financial responsibility. This research can be a useful tool in raising awareness about the importance of directed behavior for financial responsibility, particularly for policymakers.*

**165**

## 1. Introduction – the context

In recent decades, we have witnessed significant changes in the structure and main features of the financial environment, owing in part to innovative technologies, the development of the internet and various forms of online communication, as well as financial globalization, customer financial needs, and so on. The fact that lockdowns are increasing the use of digital financial services confirms the relevance and significance of the subject. For the foreseeable future, the financial landscape, as well as the banking industry, will face a variety of challenges that will have a significant impact on them. Digital financial services will help to improve economic growth by increasing financial inclusion. The future belongs to the online environment, and particularly to those who can adapt to this environment.

Based on these arguments and realities, the primary research question with which the study began was: what are the most important factors that contribute to the process of financial and digital inclusion? To address this question, we established a first goal for the paper: to determine the most crucial factors involved in the process of financial and digital inclusion. Using a holistic approach, the article proposes the big picture of the theoretical process of financial and digital inclusion. We believe that if policymakers and other stakeholders consider the entire picture of this process, based on the determining factors (of which we believe digital and financial education are the most important), a community (e.g., a country or different region such as the European Union) could be able to gain potential financial responsibility.

However, it is difficult to imagine that the increase in acceleration in adapting to the online environment would occur so abruptly, almost "overnight." This phenomenon is very well captured by physics under the name "jerk" and measured in $m\ /\ s^3$. If we assume that this name represents the pandemic that has affected many lives and turned the world upside down, we can conclude that this is the time when the online environment has become not only an option, but in many cases, the only saving solution. As a result, another research question was posed in this article: how prepared are states to make this transition, and how far have they progressed in terms of financial and digital inclusion? To address the second question, we established a new goal for the paper: determining the level of financial and digital inclusion in the research sample countries. To achieve this goal, we applied the theoretical frameworks of Analytical Hierarchy Process AHP and Fuzzy AHP management methods, as well as statistical data and critical approach information pertaining to the banking industry for the five countries selected (Bulgaria, Croatia, Czech Republic, Poland and Romania). The study can be applied to any other country of interest by using these models. The AHP (Saaty, Vargas, 2012) offers clarification on the importance of evaluation criteria, while the fuzzy approach allows for linguistic variables.

Applying AHP and Fuzzy AHP Management Methods to Assess the Level of
Financial and Digital Inclusion

The main contributions of this paper refer to the fact that we have proposed a holistic framework for evaluating the process of financial and digital inclusion, based on factors (we stress here the importance of digital and financial education) that, when effectively managed, can lead to the highest possible level of financial responsibility (a concept in the proper sense of this paper, explained here) in the case of countries or regions. Also, in this paper, we assess potential driving factors to level 3 (financial responsibility) of five countries, considering two groups of individuals who are active in the financial environment (15-34 and 35-59 age groups). We attempted to bring this element of novelty into the article by conducting a comparative analysis of the degree of financial and digital inclusion for two different age groups, allowing us to draw conclusions about the behavioral differences influenced by age in terms of digital and financial education orientation. The article proposes the AHP and Fuzzy AHP approaches as a framework for assessing the mechanism of financial and digital inclusion. As a result, an important contribution of this study is that it appears as an alternative to other studies that have conducted country hierarchies using financial and digital inclusion indexes. By applying these methods, we discovered that, based on 2017 Financial Inclusion Data, Czech Republic appears to be the closest to achieving the potential financial responsibility level for individuals aged 35 to 59. On the contrary, our findings for Croatia show that individuals between the ages of 15 and 34 are more likely to achieve a higher level of financial responsibility (according to AHP). In the case of the other countries - Bulgaria, and Romania - the hierarchy is the same regardless of the age of the citizens analyzed. Romania appears to be in the worst position when it comes to financial responsibility, and their stakeholders need to be more involved in this area.

The reminder of the paper is structured as follows: Section 2 discusses the literature on the process of financial and digital inclusion and proposes a holistic model for constructing the process of financial and digital inclusion in the global context. Section 3 introduces the methodological aspects of research and presents the AHP Management Method and the Fuzzy AHP Management Method, both of which are used to evaluate financial and digital inclusion. It also includes a presentation of the model's financial and digital inclusion factors, as well as the study's empirical findings. Section 4 summarizes the research findings and discussions, and Section 5 concludes.

## 2. Proposing a holistic model for building the global process of financial and digital inclusion – theoretical approach

According to the World Bank (*Overview*), the term "financial inclusion" has gained popularity since the early 2000s because of the identification of financial exclusion as a direct link to poverty. Financial inclusion programs, in general, aim to reach out to individuals who are unbanked or underbanked and provide them with sustainable financial opportunities (World Bank, *Global*

**167**

Bogdan Mârza, Renate Bratu, Răzvan Șerbu, Sebastian Stan, Camelia Oprean-Stan
_______________________________________________________________

*Financial Development Report 2014*). Women and disadvantaged people in rural areas were among the unbanked. Financial inclusion aims to eliminate all obstacles on both the supply and demand sides. Financial institutions are the source of supply-side barriers. They also signal a lack of financial infrastructure, such as a scarcity of financial institutions in the region, high account opening fees, or documentation requirements. Financial illiteracy, a lack of financial ability, or cultural beliefs that affect financial decisions are all factors that influence financial decisions and are examples of demand side barriers (Shankar, 2013).

Digital financial inclusion, also known as fintech-enabled financial inclusion, is the use of digital technologies to obtain access to and use structured financial systems, such as through cell phones (both smart and non-smart phones) and computers (to access the internet). Fintech firms and financial institutions both offer services under this umbrella. Financial inclusion can be facilitated by technology-enabled developments (Fernandes, 2021). Mobile money, online accounts, automatic deposits, insurance, and credit are examples of inclusive digital financial services, as are variations of these and newer financial technology (fintech) applications. Digital financial services, for example, may provide low-income families with inexpensive and easy resources to help them expand their economic opportunities. Digital finance, according to some scholar (Sahay et al., 2015), is growing financial inclusion at a time when conventional financial inclusion is decreasing. Digital financial inclusion is a vital component of efforts to include groups of people who are not part of the existing financial system. This increased financial influence has the potential to improve gender equity and economic development (Better than Cash Alliance, 2020). Digitization has enabled many people who were previously uninvolved in the financial system to benefit from financial services (Mhlanga, 2020).

In terms of metrics for measuring financial and digital inclusion through a financial inclusion index, traditional financial services have been used mostly because the emphasis has been largely on financial inclusion (Nguyen, 2020), while the position of digital finance is ambiguous. As a result, in addition to traditional financial inclusion factors reflecting the availability, usage, and affordability of financial services, as well as the users' financial literacy and ability, the factors of a comprehensive measure of financial and digital inclusion should include the availability of digital means and their use in financial services. To fill this void, we cite Shen et al. (2020), which integrated many digital elements of financial inclusion into their report and created a systematic index of digital financial inclusion for 101 countries in 2017, as well as Yang & Zhang (2020).

While, as previously stated, the process of financial and digital inclusion has long been studied in the literature, we discovered the following knowledge gap after a thorough examination: the fact that the process of digital financial inclusion has not yet been studied holistically, allowing it to be anchored in and influenced by the current context. As a result of this research, we can fill this void by proposing a comprehensive model for building a financial and digital inclusion

**168**

Applying AHP and Fuzzy AHP Management Methods to Assess the Level of
Financial and Digital Inclusion

mechanism in the global context. Another gap in the literature's information is that there is no agreed-upon and uniform method for assessing financial and digital inclusion, only partial measures. As a result, the article recommends using the AHP and Fuzzy AHP approaches as a framework for assessing the mechanism of financial and digital inclusion, in which the variables listed as the most important in inclusion are assigned different coefficients of importance. As a result, an important contribution of this study is that it appears as an alternative to other studies that have conducted country hierarchies using financial and digital inclusion indexes (we mention in this sense the "digital financial inclusion index" provided by the Internet Finance Research Center of Peking University).

Financial and digital growth, as well as financial digital access or digital financial inclusion development, have become crucial conditions of action in the future world, where people can strive to remain safe, innovate, and become wealthy (financially and non-financially).

We believe that the overall image of the theoretical phase of financial and digital inclusion could resemble that shown in Figure 1. Financial responsibility, according to the authors of this paper (Bratu, 2019), can be described as the repeated acts of using money to build long-term added value for individuals, the economy, and society. We should point out that we are assessing financial responsibility as a mechanism that can manifest itself over time and in a specific spatial sense.

General education, viewed as a continuous process of individuals becoming open minded to acquired knowledge, abilities, and skills, good health of those individuals that allows them to do different things (mentally and physically) (McKinsey Global Institute, 2020), and the existence and use of technology are, in our opinion, the main drivers of financial responsibility and financial and digital inclusion (infrastructure, instruments, specific knowledge, etc.). Thus, on education, health and technology basis we can implement and develop the process of financial and digital inclusion of individuals on financial system as a whole, concretized in financial responsibility on the top of Figure 1. At the same time, key players must continue to support the advancement of the digital economy and financial-banking environment in terms of product and service supply, organization diversity, and capital market development (policy makers, financial regulation authorities, private corporate, and households).

As shown in Figure 1, financial responsibility is built on financial education as well as digital education. In our analysis, we include banking digitization (or digital banking), which necessitates conditions such as generational cohort characteristics, education, financial assets, digital banking investments, possessions – smartphones, tablets, laptops, and computers –, energy and connectivity, and attitude (Bratu, Petria, 2018).

According to Figure 1, financial responsibility (FR) can be expressed mathematically using the following key elements:

$$FR = f(Basics, Level\ 1, Level\ 2, Context, Time) \qquad (1)$$

**169**

Bogdan Mârza, Renate Bratu, Răzvan Șerbu, Sebastian Stan, Camelia Oprean-Stan
_________________________________________________________________

As a result, we understand key elements as:
$$Basics = f\ (Education\ E, Health\ Hh, Technology\ Th) \quad (2)$$
$$Level\ 1 = f\ (Fin.Develop.FD, Bank.Develop.BD, Dig.Develop.DD) \quad (3)$$
$$Level\ 2 = f\ (Financial\ Education\ FE, Digital\ Education\ DE) \quad (4)$$
$$Level\ 3 = f\ (Financial\ Responsibility\ FR) \quad (5)$$

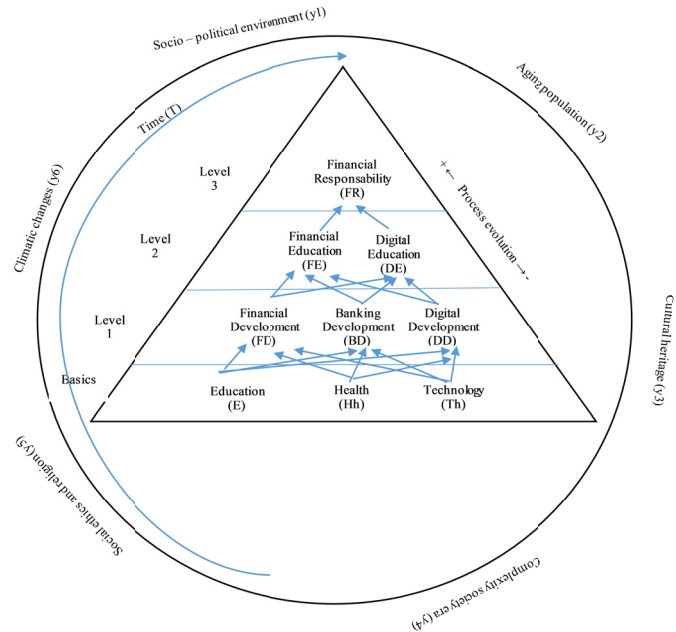

**Figure 1. Harmonized key levels of financial and digital inclusion process**

We believe that financial responsibility is attained when its components are examined in their historical context as interconnected conditions or parts in which financial responsibility can manifest. In our opinion, at least six quality variables from $y_1$ to $y_6$ shown in Figure 1 have a strong influence on context (where $y_1$ = socio-political environment; $y_2$ = aging population; $y_3$ = cultural heritage; $y_4$ = complexity society era; $y_5$ = social ethics and religion; $y_6$ = climatic changes).

As a result, we have a mathematical expression for context:
$$Context = f(y_1, y_2, y_3, y_4, y_5, y_6) \quad (6)$$

Finally, we believe that if these components are managed and developed optimally, a community (e.g., a country or different region such as the European Union) may be able to gain potential financial responsibility. Policymakers and other stakeholders must take a holistic approach to this concept, as expressed in the following expressions:

$$FR = f\left[(E, Hh, Th) \to (FD, BD, DD) \to (FE, DE)\right] : f\left[(y_1, y_2, y_3, y_4, y_5, y_6), T\right] \quad (7)$$

We propose that policymakers and other stakeholders create sets of relevant indicators for each key component and level of the financial responsibility process to evaluate the process's positive evolution (, +). We believe that the most important indicators are those reported at the international level and determined using comparable methods. For example, in the case of Education (E), PISA results could be used; in the case of Digital Development (DD), the European Commission's DESI INDEX; in the case of Financial Education (FE), a set of Financial Inclusion Indicators published by the G20; and so on.

### 3. Methodological Aspects of Research
### 3.1. Research design

In this study, we propose using the Analytic Hierarchy Process (AHP) Management Method as an evaluation methodology to analyze the level of financial and digital inclusion for the sample countries, and we use the Fuzzy Analytic Hierarchy Process (Fuzzy AHP) Management Method to confirm the obtained results (test the robustness of the results). The factors identified as the most important in the process of financial and digital inclusion, developed according to the holistic model proposed in the global context (and represented in Figure 1), are assigned different importance coefficients in these methods.

**Analytic Hierarchy Process (AHP)**

AHP is a method for organizing, assessing, and reviewing complex decisions. Thomas L. Saaty developed AHP in the 1970s, and it has been widely researched and perfected since then. It is based on mathematics and psychology and has been extensively researched for the last 20 years due to its broad applicability and ease of use, and as a result, the academic literature in the field has become rich and comprehensive (Păunescu & Moraru, 2018; Saaty, Vargas, 2012; Hasan & Rahman, 2017). It is particularly useful in group decision-making and is used in a broad range of decision-making situations around the world, including government, enterprise, industry, healthcare, and education. AHP has been used for linear / integer / target programming, data envelope processing, genetic algorithms, balanced score cards, and neurol networks (Millet & Wedley, 2002). AHP is also used in evaluations in addition to being a commonly used decision-making method. It is used to evaluate financial service efficiency (Turskis, 2011), financial performance for wealth management (WM) banks (Tsai, 2016), and higher education teaching performance (Thanassoulis, 2017).

The AHP implementation procedure in this article is as follows, and it consists of the following main steps for assessing the level of financial and digital inclusion.

**171**

Bogdan Mârza, Renate Bratu, Răzvan Șerbu, Sebastian Stan, Camelia Oprean-Stan
______________________________________________________________________

**Step 1:** Build the hierarchy from the top down, starting with the financial responsibility goal and then with the criteria from a broad perspective (Figure 1).

**Step 2:** Create a matrix of pairwise comparisons of sub-criteria:

$$A = \begin{bmatrix} a_{11} & a_{12} & \cdots & a_{1n} \\ a_{21} & a_{22} & \cdots & a_{2n} \\ \vdots & \vdots & \ddots & \vdots \\ a_{n1} & a_{n2} & \cdots & a_{nn} \end{bmatrix} \qquad (8)$$

Selected factors (see Table 1) were combined in pairs and inserted into the evaluation matrix. Experts use the AHP method to compare sub-criteria by filling out a pairwise comparison matrix.

$$A = (a_{ij})_{n \times n} \qquad (9)$$

where: $\quad a_{ji} = \frac{1}{a_{ij}}$, i, j = 1, 2, ..., n $\qquad (10)$

$$a_{ii} = a_{jj} = 1 \qquad (11)$$

Experts recommended using a nine-point scale to complete individual comparative matrixes, where "1" means that factors are similarly important and "9" means that one factor is highly important over another.

**Step 3:** A widely used method to estimate the priority vector of the form $w = (w_1, \cdots, w_n)^T$ (where n represents the number of sub-criteria) is the geometric mean method, proposed by Crawford and Williams (1985):

$$w = \left(\prod_{j=1}^{n} a_{ij}\right)^{1/n} \times \left[\sum_{i=1}^{n}\left(\prod_{j=1}^{n} a_{ij}\right)^{1/n}\right]^{-1} \qquad (12)$$

**Step 4:** The Geometric consistency index (GCI), developed by Crawford and Williams (1985) and refined by Aguarón (2003), is calculated to examine the inconsistency of the pair-wise matrix when the priority vector $w = (w_1, \cdots, w_n)^T$ was estimated using the geometric mean method.

With the weights calculated in the previous step, a local quantification of inconsistency $e_{ij}$ can be constructed for each entry $a_{ij}$:

$$e_{ij} = a_{ij}\frac{w_j}{w_i}, i, j = 1, \ldots, n. \qquad (13)$$

and it is calculated as follows:

$$GCI = \frac{2}{(n-1)(n-2)} \sum_{i=1}^{n-1} \sum_{j=i+1}^{n} (\ln e_{ij})^2 \qquad (14)$$

The value threshold for this indicator is determined by the number of sub-criteria that are compared (n). In our study n>4, resulting in a $GCI \leq 0,37$.

**Step 5:** At this point, Saaty proposed creating a separate matrix for each criterion. As a result, a matrix $A^{(k)}$ is the matrix of pairwise comparisons between alternatives according to sub-criteria k. Then, we estimate their priority vectors $w^{(k)}$ and $GCI^{(k)}$.

**Step 6:** The specific score of alternatives (a) will be calculated to assess the final order by adding the weights for alternatives multiplied by the weights of the corresponding sub-criteria, using the following formula:

$$S_a = \sum_{i=1}^{n} w_i \times w_i^{(a)} \qquad (15)$$

Applying AHP and Fuzzy AHP Management Methods to Assess the Level of
Financial and Digital Inclusion

**Fuzzy Analytic Hierarchy Process (FAHP)**

The authors of the paper used the Fuzzy AHP method to validate their findings. The explanation for this is that the AHP approach is often criticized for failing to deal with complexity and imprecision that occur when solving multi-criteria analysis problems. In line with other research articles (Kahraman et al., 2003; Amile, 2013; Ban, 2020) a Fuzzy AHP approach was used here because it eliminates this flaw, making the process more flexible while retaining accuracy. Another reason to use the Fuzzy AHP approach is that it has been used in previous studies for evaluations, such as determining the performance of businesses using financial and non-financial metrics (Ban, 2020).

The following is a summary of the technique used in this article in the Fuzzy AHP method.

**Step 1.** Linguistic terms are used to express pairwise comparisons in Fuzzy AHP. The linguistic terms are represented by membership functions, which are typically triangular. We can write the fuzzy pair-wise comparison matrix for the chosen sub-criteria using the triangular fuzzy scale formed by the triangular fuzzy number of the form $(l, m, u), where\ l \leq m \leq u$:

$$\tilde{A} = (\tilde{a}_{ij})$$

where

$$\begin{cases} \tilde{a}_{ij} = (l_{ij}, m_{ij}, u_{ij}),\ i,j = 1, \ldots, n\ and\ i < j \\ \tilde{a}_{ji} = \left(1/u_{ij}, 1/m_{ij}, 1/l_{ij}\right),\ i,j = 1, \ldots, n\ and\ i > j \\ \tilde{a}_{ii} = \tilde{a}_{jj} = (1,1,1),\ i,j = 1, \ldots, n\ and\ i = j \end{cases} \quad (16)$$

If $A = (a_{ij})$ is a positive reciprocal matrix, Buckley (1985) discovered that $\tilde{A} = (\tilde{a}_{ij})$ is a fuzzy positive reciprocal matrix. That is, if the outcome of $A = (a_{ij})$ comparisons is consistent, it can be assumed that the result of $\tilde{A} = (\tilde{a}_{ij})$ comparisons is also consistent.

**Step 2.** The value of fuzzy weights $\tilde{w}$ with respect to the $i$-th sub-criteria is defined using the geometric mean for a fuzzy number as:

$$\tilde{w} = \left(\prod_{j=1}^{n} \tilde{a}_{ij}\right)^{1/n} \otimes \left[\sum_{i=1}^{n} \left(\prod_{j=1}^{n} \tilde{a}_{ij}\right)^{1/n}\right]^{-1} \quad (17)$$

**Step 3:** Finally, weights are calculated for each criterion through defuzzification (Shaverdi, 2012) and normalization ($\tilde{w}_i$ is of the form $(l_i, m_i, u_i)$, and $W_i$ is a crisp number):

$$W_i = \frac{(u_i - l_i) + (m_i - l_i)}{3} + l_i = \frac{l_i + m_i + u_i}{3} \quad (18)$$

$$W_{norm_i} = \frac{W_i}{\sum_{k=1}^{n} W_k} \quad (19)$$

**Step 4:** Repeat steps 1-3 to compute the weights of the alternatives ($a$) within each sub-criteria ($W_{norm_i}^a$).

Bogdan Mârza, Renate Bratu, Răzvan Șerbu, Sebastian Stan, Camelia Oprean-Stan
_________________________________________________________________

**Step 5:** The specific score of alternatives (*a*) will be calculated to assess the final order by adding the weights for alternatives multiplied by the weights of the corresponding sub-criteria, using the following formula:

$$S_a = \sum_{i=1}^{n} W_{norm_i} \times W_{norm_i}^a \qquad (20)$$

### 3.2. The Model's Presentation of Sub-Criteria Indicators

The article is based on an examination of some indicators published in the World Bank Database and the G20 Financial Inclusion Indicators, specifically referring to the most important aspects of the Financial and Digital Inclusion processes, and it refers to the sub-criteria indicators that characterize the two main criteria of responsibility: Digital education and development of banking sector environment (Digital inclusion selection: Go digital – 2017 indicators) and Financial education and development, that refers to specific infrastructure which facilitated customers to acquire banking products and services (Financial inclusion selection: Use of accounts – 2017 indicators).

A review of the literature (Kelikume, 2021; Shen et al., 2020; Yang & Zhang, 2020) yielded 11 evaluation indices related to the financial and digital inclusion processes, which were summarized. "DE: Digital Education ($i_1$–$i_5$)" and "FD: Financial Education ($i_6$–$i_{11}$)" are the two criteria. The importance levels of the criteria in comparison to each other are determined through expert consultation and converted to numerical values using the scale of relative importance.

As shown in Table 1, we believe that digital education is a more important criterion than financial education in determining financial and digital inclusion. The argument is that we now live in a time when technology is an integral part of our lives. We are constantly using technology to access information, develop abilities and skills, and carry out professional activities. We are now living in an overly complex digital society era, which will only become more complex in the coming years. Consumers have migrated to the electronic environment and spend significantly more time online; as a result, an increasing number of institutions and businesses have significantly increased their work in this direction. Online shopping is becoming increasingly popular in the European Union and around the world. Along with the increased use of the Internet and the improvement of security standards, consumers value the ability to shop whenever and wherever they want, with access to a wide range of online financial services. The financial market is expected to face new challenges during the current pandemic conditions, with restricted travel and consumers under social distancing restrictions, and its digitized instruments to continue to develop despite the effects of the economic crisis.

Another argument that explains why digital development is more important than financial development is that a new environmental context necessitates a new transformation of the physical "image" of bank networks. We argue once more that, in the long run, we will see an increase in digital banking, so physical development of financial access infrastructure through ATMs and commercial bank branches appears less important in this context.

**Table 1. Analytical Hierarchical Framework and the Importance of Criteria**

| Level 1: Goal | Level 2: Criteria | Importance | Level 3: Sub-criteria | Group Importance | Final Importance | Level 4: Alternative |
|---|---|---|---|---|---|---|
| FINANCIAL RESPONSIBILITY | Digital Education | 2 | ($i_1$) Access to a mobile phone (% ages 15-34 and % ages 35-59) | 1 | 2 | Bulgaria Croatia Czech Republic Poland Romania |
| | | | ($i_2$) Access to internet (% ages 15-34 and % ages 35-59) | 1 | 2 | |
| | | | ($i_3$) Made or received digital payments in the past year (% ages 15-34 and % ages 35-59) | 2 | 4 | |
| | | | ($i_4$) Made payment using a mobile phone or the internet (% ages 15-34 and % ages 35-59) | 2 | 4 | |
| | | | ($i_5$) Used a mobile phone or the internet to check account balance in the past year (% ages 15-34 and % ages 35-59) | 2 | 4 | |
| | Financial Education | 1 | ($i_6$) ATMs per 100,000.00 adults | 1 | 1 | |
| | | | ($i_7$) Commercial Bank Branches per 100,000.00 adults | 1 | 1 | |
| | | | ($i_8$) Active account (% ages 15-34 and % ages 35-59) | 2 | 2 | |
| | | | ($i_9$) Borrowed from a financial institution or used a credit card (% ages 15-34 and % ages 35-59) | 3 | 3 | |
| | | | ($i_{10}$) Saved at a financial institution (% ages 15-34 and % ages 35-59) | 4 | 4 | |
| | | | ($i_{11}$) Saved for old age (% ages 15-34 and % ages 35-59) | 5 | 5 | |

### 3.3. Study sampling and data collection

For the purposes of this article, we used the theoretical framework of the AHP management method, as well as statistical data and critical approach information pertaining to the banking industry in the five countries chosen (Bulgaria, Croatia, Czech Republic, Poland, and Romania). The study can be applied to any other country of interest by using this model. We chose these five ECE countries primarily because they share a common interest. After the communist era, their economies had a lot in common. They were barely affected by communism and liberated themselves in the same period 1998-1990, launching their free market economy around the same time under similar conditions, struggling until then to gain trust and appreciation for their economy in general and the financial market.

The World Bank's Global Findex database (https://globalfindex.worldbank.org/) was the primary source for collecting indicators on financial inclusion at the country level, as it is the most detailed data collection on how adults invest, borrow, pay bills, and handle risk in the country. Since 2011, the database has been published every three years. The G20 Financial Inclusion Indicators developed by the Global Partnership for Financial Inclusion (GPFI) is a widely used data portal for evaluating traditional financial inclusion, and because the portal added new indicators measuring usage of digital payments and access to digital infrastructure in 2016, it has also become an important source that can provide the characteristics of digital finance.

This is the argument supported by an examination in this article of some indicators published in the World Bank Database and the G20 Financial Inclusion Indicators, with a focus on the most critical aspects of the financial and digital inclusion processes. The data for the five countries studied are from 2017.

### 4. Research Results and discussions
#### 4.1. Empirical evidence of the use of the AHP Method in the financial and digital inclusion process

The values of the factors (age of respondents 13-34 years) for each country were considered for the multi-criteria analysis performed in this paper using the AHP method (Geometric mean). We created a pairwise comparison matrix using the scale of relative importance and calculated normalized weights of factors using geometric mean, starting with the importance assigned to the criteria.

The Geometric consistency index GCI = 0.1189 is used to evaluate the consistency of the pairwise comparison matrix. The results show a value less than 0.37 (for n = 11), indicating that the matrix under consideration is consistent.

We calculated the weight of each country using the AHP method (geometric mean) after determining the weights of the factors, as well as the Geometric consistency rate (GCI) within the sub-criteria $i_1$, *Access to a mobile phone*. We followed the same procedure for each of the analyzed factors. Using the AHP method, the score for each country is calculated by summing the weighted values, and the final ranking is established. We proceeded in the same manner for the respondents' age group of 35-59 years. Using the AHP method, the score for each country is calculated by summing the weighted values, and the final ranking is established.

#### 4.2. Model validation using the Fuzzy AHP Method in the financial and digital inclusion process

The values of the factors (age of individuals 15-34 years) for each country were considered for the multicriteria analysis using the Fuzzy AHP method presented in this paper.

**176**

Applying AHP and Fuzzy AHP Management Methods to Assess the Level of
Financial and Digital Inclusion

**Table 2. Fuzzy triangular scale**

| | |
|---|---|
| (1,1,1) | equal importance |
| (1,2,3) | intermediate values |
| (2,3,4) | moderate importance |
| (3,4,5) | intermediate values |
| (4,5,6) | strong importance |
| (5,6,7) | intermediate values |
| (6,7,8) | very strong importance |
| (7,8,9) | intermediate values |
| (8,9,9) | extreme importance |
| (1/3,1/2,1); (1/4,1/3,1/2); (1/5,1/4,1/3); (1/6,1/5,1/4); (1/7,1/6,1/5); (1/8,1/7,1/6); (1/9,1/8,1/7); (1/9,1/9,1/8) | reciprocal scale |

Starting with the criteria's importance (Table 1), we created a pairwise comparison matrix using a fuzzy triangular scale (Table 2), and we calculated normalized weights of factors using geometric mean.

After determining the weights of the factors, we used the Fuzzy AHP method to calculate the weight of each country within the factor $i_1$. We followed the same procedure for each of the analyzed factors. The score for each country was determined by adding the weighted values and determining the final ranking.

We followed the same procedure for the respondents' age group of 35-59 years. Each country received a score.

### 4.3. Findings

Starting with the initial data, the rankings with the countries are obtained based on the parameters considered, taking into account those two groups of individuals (ages 15-34 and 35-59). We emphasize once more that the number of parameters used for analysis is eleven, and it is not possible to make an accurate estimation of country rankings without using a complex hierarchy method. As a result, we chose the AHP Method, and for validating the results, we used the Fuzzy AHP Method. As can be seen, there are no significant differences between countries in terms of obtained results. Two countries (Romania and Bulgaria) have the lowest levels of financial and digital inclusion, while the Czech Republic, Croatia, and Poland have the highest levels. Furthermore, by correlating the conceptual framework presented in Figure 1 with mathematical expression no 7 and the results of AHP and Fuzzy AHP Method, we can draw a relative position of countries in alignment with the financial responsibility process (Table 3).

We notice that when the two methods, AHP and Fuzzy AHP, are used, the results are similar, which confirms and verifies the results obtained. According to

Bogdan Mârza, Renate Bratu, Răzvan Șerbu, Sebastian Stan, Camelia Oprean-Stan
_________________________________________________________________

the AHP method, the first group of people - those aged 15 to 34 - rank the analysed countries in the following order: Croatia is first, followed by the Czech Republic, Poland, Bulgaria, and Romania, while the only difference according to the Fuzzy AHP method is that the Czech Republic is first, and Croatia is second. That is, individuals in the 15-34 age group are more financially included in Croatia (according to AHP) and the Czech Republic (according to Fuzzy AHP) than in the other countries. The results obtained for the second group of individuals aged 35-59 are identical for both methods and differ significantly from the other age group in that the countries are ranked as follows: Czech Republic, Poland, Croatia, Bulgaria, and Romania. Individuals in the Czech Republic in the 35-59 age group appear to be the most financially included when compared to the same group of adults in Poland, Croatia, Bulgaria, and Romania, which is ranked last among the ECE countries.

**Table 3. Financial and Digital Inclusion Indicators – AHP Method and Fuzzy AHP results, Final ranking**

| | AHP Method |
|---|---|
| 15-34 ages: | Romania (the most distanced of Level 3), Bulgaria, Poland, Czech Republic, Croatia (the closest to reach Level 3) |
| 35-59 ages: | Romania (the most distanced of Level 3), Bulgaria, Croatia, Poland, Czech Republic (the closest to reach Level 3) |
| | **Fuzzy AHP Method** |
| 15-34 ages: | Romania (the most distanced of Level 3), Bulgaria, Poland, Croatia, Czech Republic (the closest to reach Level 3) |
| 35-59 ages: | Romania (the most distanced of Level 3), Bulgaria, Croatia, Poland, Czech Republic (the closest to reach Level 3) |

### 5. Conclusions

The primary goal of this paper was to develop a framework for the financial responsibility process and to investigate it using financial and digital inclusion data that we gathered for this paper. Another goal of the paper was to assess the level of financial and digital inclusion in the study sample countries. Financial data that we process are addressed to five ECE countries that we specifically chose because they are in the same area of interest and have a lot in common (especially because they were liberated from a hardly communist regime at the same time: Bulgaria, Croatia, Czech Republic, Poland and Romania). To accomplish this, we used the theoretical frameworks of Analytical Hierarchy Process AHP and Fuzzy AHP management methods, as well as statistical data and critical approach information from the banking industry in the five countries chosen.

We assess a portion of the potential driving to level 3 (financial responsibility) of those 5 selected countries in this paper, considering two groups

Applying AHP and Fuzzy AHP Management Methods to Assess the Level of
Financial and Digital Inclusion

of individuals who are active in the financial environment (15-34 and 35-59 age groups). Using the AHP Method, we discovered that, based on 2017 Financial Inclusion Data, the Czech Republic appears to be the closest to achieving the potential financial responsibility level for individuals aged 35 to 59. In contrast, our findings for Croatia show that individuals between the ages of 15 and 34 are more likely to achieve a higher level of financial responsibility. In the case of the other countries – Bulgaria and Romania – the hierarchy is the same, regardless of the age of the citizens analyzed. Using the Fuzzy AHP Method, we discovered that, based on 2017 Financial Inclusion Data, the Czech Republic appears to be the closest to achieving the potential level of financial responsibility for all individuals in both age groups. If this method is used, Romania appears to be in the worst position in terms of financial responsibility, and their stakeholders need to be more involved in this area.

We attempted to add a novel element to the article by conducting a comparative analysis of the degree of financial and digital inclusion for two different age groups, allowing us to draw conclusions about behavioral differences influenced by age in terms of orientation toward digital and financial education. Among other things, the findings show that while Polish people aged 15-34 outperform Croatians of the same age group in terms of financial and digital inclusion, the situation shifts in the other age group (35-59), when Poles outnumber Croatians. These age differences are quite normal when one considers an individual's changing status - from student to professional, gaining constant experience as a professional and as a consumer of financial-banking products and services, as well as different fluctuations in one's own income. There was a distinction between countries and age groups. According to the findings, Romania reported the lowest level of indicators for both age groups, and as a result, Romania was ranked last in terms of financial and digital inclusion data. In this case, we believe that the main cause of those lower levels is the lower level of financial inclusion indicator of Romanians, which is driven by the lowest level of financial education process. In fact, Romania is ranked last among EU countries (22 percent of Romanians reached a certain level of financial literacy - EU 2015) and 123rd out of 143 countries globally. A few years ago, approximately 40% of adults in Romania were unbanked (the EU average is 10%), compared to adults in Austria or Germany, where the same indicator is less than 1% (Romanian Association of Banks, 2020). As a result, we strongly believe that the financial literacy process is critical in terms of knowledge, abilities, and skills for individuals (at various stages of age and professional experience) (Panos & Wilson, 2020).

Along with a lack of financial literacy, we believe that another important factor contributing to this situation is the level of general education (formal and informal), which serves as the foundation for other types of specialized education such as financial or digital edification. In this section, we will discuss data from OECD reports on PISA results from 2015 and 2018. However, we believe that

**179**

Bogdan Mârza, Renate Bratu, Răzvan Șerbu, Sebastian Stan, Camelia Oprean-Stan
_________________________________________________________________

higher PISA scores will determine higher levels of specialized education (financial education, digital education, health education, and so on). As a result, PISA results could be a valuable motivator for policymakers to develop and implement long-term strategies to improve general education outcomes, and then to stimulate the improvement of financial and digital edification. Looking specifically at the 2018 PISA results, Bulgaria and Romania were classified as having less than 450 points (426.7 and 428 points, respectively), Croatia and the Czech Republic as having 450-500 points (471.7 and 495.3 points, respectively), and Poland as having more than 500 points (11th place with 513 points) (Facts Maps. PISA 2018 Worldwide Ranking).

To support the direct correlation between education and specialized education, we note that Poland and Bulgaria also took part in another OECD survey relating to the question – are students financially savvy? Poland received 520 points in this category (which is higher than the OECD average of 505 points) and Bulgaria received 432 points (being significantly below the OECD average). The main factors influencing the ability of future adults - 15-year-old students - to make proper financial decisions in terms of benefits and risks are demographic area type, parents' level of education, family wealth (income and possessions), immigrant background, gender, and so on. Some of those drivers we included in what is called ESCS - PISA Index of Economic, Social and Cultural Status (OECD PISA Results, 2018). In this context, policymakers and other stakeholders must be aware of their role in the implementation of national financial and digital education strategies.

We believe that the most important idea of this research is that it can be a useful tool, particularly for policymakers, in raising awareness about the importance of directed behavior for financial responsibility. They can also develop a country-specific strategy for specific groups of people (categorized by age, education, or income level, or industry in which they work) to encourage them to become more involved in a more digital financial environment (using specific financial products and services) and to use financing, investing, and philanthropy to reduce or solve major humanitarian issues.